\documentclass[aps,pre,address,twocolumn]{revtex4}
\usepackage[english]{babel}

\usepackage{color}

\usepackage{epsfig}
\usepackage{graphicx}
\usepackage{epstopdf} 
\usepackage{bm}
\usepackage{times}
\usepackage{amsfonts}
\usepackage{amssymb}
\usepackage{amsmath}
\usepackage{cleveref}
\usepackage{color}
\usepackage{wasysym}

\usepackage{algorithmicx}

\begin{document}
\title{Kardar-Parisi-Zhang Universality of the Nagel-Schreckenberg Model}

\author{Jan de Gier$^{1}$, Andreas Schadschneider$^2$, Johannes
  Schmidt$^{2,3}$, Gunter M.~Sch\"utz$^{4}$ } \affiliation{$^{1}$ARC
  Centre of Excellence for Mathematical and Statistical Frontiers
  (ACEMS), School of Mathematics and Statistics, The University 
  of Melbourne, VIC 3010, Australia\\
  $^{2}$Institut f\"{u}r Theoretische Physik, Universit\"{a}t zu
  K\"{o}ln, 50937 Cologne, Germany\\
  $^{3}$Bonacci GmbH, Robert-Koch-Str. 8, 50937 Cologne, Germany \\
  $^4$ Theoretical Soft Matter and Biophysics, Institute of Complex
  Systems II, Forschungszentrum J\"ulich, 52425 J\"ulich, Germany
}

\begin{abstract}
  Dynamical universality classes are distinguished by their dynamical
  exponent $z$ and unique scaling functions encoding space-time
  asymmetry for, e.g. slow-relaxation modes or the distribution of
  time-integrated currents.  So far the universality class of the
  Nagel-Schreckenberg (NaSch) model, which is a paradigmatic model for
  traffic flow on highways, was not known except for the special case
  $v_{\text{max}}=1$. Here the model corresponds to the TASEP (totally
  asymmetric simple exclusion process) that is known to belong to the
  superdiffusive Kardar-Parisi-Zhang (KPZ) class with $z=3/2$.  In
  this paper, we show that the NaSch model also belongs to the KPZ
  class \cite{KPZ} for general maximum velocities
  $v_{\text{max}}>1$. Using nonlinear fluctuating hydrodynamics theory
  we calculate the nonuniversal coefficients, fixing the exact
  asymptotic solutions for the dynamical structure function and the
  distribution of time-integrated currents. Performing large-scale
  Monte-Carlo simulations we show that the simulation results match
  the exact asymptotic KPZ solutions without any fitting parameter
  left. Additionally, we find that nonuniversal early-time effects or
  the choice of initial conditions might have a strong impact on the
  numerical determination of the dynamical exponent and therefore lead
  to inconclusive results.  We also show that the universality class
  is not changed by extending the model to a two-lane NaSch model with
  dynamical lane changing rules.

\end{abstract}

\date{\today }
\maketitle

\section{Introduction}

In statistical physics, nonequilibrium systems are divided into
universality classes according to their dynamical behavior. Dynamical
universality classes are distinguished by their dynamical exponent $z$
and unique scaling functions encoding space-time asymmetry for,
e.g. slow-relaxation modes or the distribution of time-integrated
currents.  Systems in the same universality class show for large times
of order $1\ll t \ll L^z$, where $L$ is the length of the system,
identical statistical properties, while local interactions are coded
in nonuniversal scaling factors. The two most prominent examples are
the diffusive class with dynamical exponent $z=2$ and the
superdiffusive Kardar-Parisi-Zhang (KPZ) class with $z=3/2$
\cite{KPZ,Halp15}.  A generic example for the latter is the totally
asymmetric simple exclusion process (TASEP)
\cite{asep,Derrida,Schuetz00} which describes the single-file motion
of uni-directionally moving particles on a discrete one-dimensional
lattice. Due to exclusion each lattice site can accommodate at most
one particle.

The first indirect proof of KPZ-universality in the TASEP and its
partially asymmetric generalization (ASEP) came from finite-size
scaling analysis of the spectral gap of the Markov generator, using
the Bethe ansatz \cite{Dhar,Gwa92,Kim95}.  These results yield the
dynamical exponent $z=3/2$. Since then there has been remarkable
progress that has led to a much more detailed understanding of the
fluctuations in one dimensional systems using techniques from random
matrix theory \cite{FP,Spohn06,S07,Corwin2012,QS2015}. Several exact
solutions for models in the KPZ universality class, such as the ASEP
and the KPZ equation \cite{Johansson2000,BR, Prae02, Prae04,FS,
  TW2009a, SS2010,ACQ2011}, have resulted in explicit expressions for
universal distribution functions and correlations of physical
quantities in appropriate scaling limits.

For applications to highway traffic the TASEP has been generalized to
the Nagel-Schreckenberg (NaSch) model \cite{NaSch92}. Here the
particles have an internal degree of freedom, called velocity, which
determines their hopping range. The velocity changes dynamically and
is limited by a maximum value $v_{\text{max}}$. In contrast to the
standard TASEP the NaSch model is defined by a parallel updating scheme
which leads to more realistic results. For the special case
$v_{\text{max}}=1$ the NaSch model reduces to the TASEP with parallel
dynamics. 
 
For more than 20 years one has tried to determine the universality
class of the NaSch model. For the case $v_{\text{max}}=1$ it was
expected that it also belongs to the KPZ class, i.e. that the use of
parallel dynamics does not change the universality class. This was
confirmed with random matrix theory in \cite{Johansson2000} and was
subsequently generalized to other parallel update schemes with
$v_{\text{max}}=1$, using determinantal techniques derived from Bethe
ansatz \cite{Rako05}.  However, for general maximum velocities
$v_{\text{max}}>1$ the universality class has remained under debate
since the internal degree of freedom (i.e. the velocity) might lead to
a different universality class and numerical studies were inconclusive
\cite{CsanyiK95,SasvariK97}.

Here we will show that the NaSch model indeed belongs to the KPZ class
for all parameter values. Using nonlinear fluctuation hydrodynamics
theory we calculate the nonuniversal coefficients fixing the exact
asymptotic solutions for the dynamical structure function and the
distribution of time-integrated currents. Performing large-scale
Monte-Carlo simulations we show that the simulation results match the
exact asymptotic KPZ solutions without any fitting parameter
left. Additionally, we find that nonuniversal early-time
effects, or the choice of initial conditions might have an
strong impact on the numerical dynamical exponent determination and
therefore lead to inconclusive results.  We also show that the
universality class is not changed by extending the model to a two-lane
NaSch model with dynamical lane changing rules. This implies that
neither the use of random-sequential dynamics nor single-file
behaviour are essential for the universality.

\section{Nagel-Schreckenberg Model}

The model introduced by Nagel and Schreckenberg (NaSch) \cite{NaSch92}
is by now regarded as a minimal cellular automation model for traffic
flow on highways. It can be viewed as an extension of the TASEP
with parallel dynamics to longer-range interactions. However, in
contrast to other generalisations of the TASEP which allow for the
movement of particles beyond the nearest-neighbour, the NaSch model
has a kind of velocity memory controlled by an internal parameter
$v$. $v_n(t)$ corresponds to the number of cells particle $n$ has
moved forward in time step $t$.  By the dynamical rules of the model,
$v_n(t)$ can at most increase by one in the next timestep which
mimics the limited acceleration properties of vehicles. This velocity
makes the model at the same time more realistic, but also more
difficult to analyse. Especially it is not clear whether it has an
impact on the dynamical universality class of the model.

The dynamical rules for the NaSch model are given by four steps which
are applied to all vehicles at the same time (parallel or synchronous
update). The update rule for the $n$-th vehicle is:

\begin{enumerate}
\item Acceleration: If $v_{n}<v_{\text{max}}$, the speed of the $n$-th
  vehicle is increased by one, but $v_{n}$ remains unaltered if
  $v_{n}=v_{\text{max}}$, i.e
\[
v_{n}\rightarrow\text{min}\left(v_{n}+1,v_{\text{max}}\right)
\]

\item Deceleration: If $v_{n}>d_{n}$, ($d_{n}$ is the headway of the $n$-th
vehicle) the speed of the $n$-th vehicle is reduced to $d_{n}$,
i.e.
\[
v_{n}\rightarrow\text{min}\left(v_{n},d_{n}\right)
\]

\item Randomization: If $v_{n}>0$, the speed of the $n$-th vehicle is
decreased randomly by unity with probability $p_s$, i.e.
\[
v_{n}\overset{p_s}{\rightarrow}\text{max}\left(v_{n}-1,0\right)\,.
\]
With probability $1-p_s$ the velocity of the vehicle remains unchanged.
The velocity $v_{n}$ does not change if $v_{n}=0$.

\item Vehicle movement: Each vehicle is moved forward according to its new
velocity determined in 1.-3., i.e.
\[
x_{n}\rightarrow x_{n}+v_{n}
\]
\end{enumerate}

These rules are minimal in the sense that the basic features of real
highway traffic (e.g. spontaneous jam formation) are no longer
reproduced if one rule is left out.  Also the order of the rules is
essential for realistic behaviour. Throughout this paper, numerical
results will refer to an implementation of the NaSch model on a
lattice with periodic boundary conditions.

\begin{figure}[h]
\centerline{\includegraphics[width=0.45\textwidth]{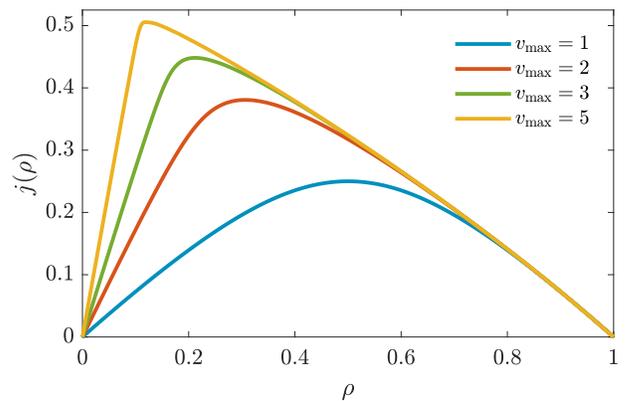}}
\caption{NaSch
  current-density plot for a stationary system with $p_s=0.25$ and
  various $v_\text{max}$. The current-density relation was estimated
  by running Monte Carlo simulations using systems of length $L=10.000$ and periodic boundary
  conditions. Finite size and statistical errors are in order of line
  width.}
\label{fig:RHO_Current_p_025}
\end{figure}

In contrast to the TASEP, so far no closed solution for the stationary
state of the NaSch model is known. The reviews
\cite{SCNBook,ChowdhurySS00,MaerivoetM05} give an overview over 
known results.
Fig.~\ref{fig:RHO_Current_p_025} shows the fundamental diagram, i.e.
the density-dependence of the stationary current, for different
values of $v_{\text{max}}$. For $v_{\text{max}}>1$ the particle-hole
symmetry of the TASEP is lost and as a result the function $j(\rho)$
is no longer symmetric around $\rho=1/2$.

\section{Nonlinear Fluctuating Hydrodynamics (NLFH)}
\label{sec:NLFH}

Nonlinear fluctuating hydrodynamics \cite{Mori,Swift,Das} is a
powerful phenomenological tool to describe the large-scale behaviour
behaviour of fluctuations of conserved quantities in many-body system
both in and out of thermal equilibrium \cite{Spoh14}. Notably, it
captures the large-scale properties of the dynamical structure
function in the universality class of the KPZ equation \cite{Halp15}
as well as an infinite discrete family of other universality classes
\cite{Popk15b,Popk16}.  In particular, the theory allows for adopting
exact results obtained for specific models \cite{Prae02,Prae04,Bern16}
to models that are not exactly solvable but are within the same
universality class.

In the following we use the non-linear fluctuating hydrodynamic
equation for conservative driven diffusive lattice gases with one
conserved density and thus establish the connection of the KPZ
equation with the NaSch model. Significantly, we will present exact
analytic predictions for the dynamical structure function and time
integrated current distributions which will serve as a test of the
KPZ-universality.

Particle conservation along with local stationarity and slow relaxation of the conserved modes
implies that the long-time evolution of the
NaSch model at large scales is described in terms of a conservation
law $\partial_t\varrho(x,t)+\partial_x\mathtt{j}(x,t)=0$, where
$\varrho(x,t)$ is the coarse-grained local density field, and
$\mathtt{j}(x,t)$ is the associated current. 
Local stationarity ensures that the current $\mathtt{j}(x,t)$ depends on $x$ and
$t$ only through the 
density $\varrho(x,t)$, i.e.,
one has $\mathtt{j}(x,t)=j(\varrho(x,t))$ with the stationary
current-density relation $j(\varrho)$ \cite{Kipn99}. Thus
$\partial_x \mathtt{j} = j'(\varrho) \partial_x \varrho$ where
$j'(\varrho) = \mathbf{d}j(\varrho)/(\mathbf{d}\varrho) $. Evidently,
$\varrho(x,t) = \rho$ with any constant $\rho$ in the physically
permissible range is the stationary solution to this hydrodynamic
equation. In this deterministic hydrodynamic description the effects
of the noise disappear because of the spatial coarse-graining of the
density and the Eulerian scaling of time in which the microscopic
space and time scales are rescaled proportionally to a common scaling
factor. In the case of lattice gas models these microscopic scales are
the lattice constant and the time scale of particle jumps between
lattice sites.

As next step one subtracts from the local density
field $\varrho(x,t)$ its stationary background $\rho$ to obtain the
fluctuation field $\mathtt{u}(x,t) = \varrho(x,t)-\rho$, and expands
the current $j(\varrho(x,t))$ in $\mathtt{u}(x,t)$ around the constant
$\rho$. 
To incorporate fluctuations and thus capture the effects of noise 
arising from the stochastic dynamics and to arrive at the
fluctuating hydrodynamic description, a phenomenological diffusion term
$\mathcal{D}\partial_x \mathtt{u}(x,t)$ and Gaussian white noise
$\mathcal{B} \xi(x,t)$ are added to the current. To capture the
universal behavior correctly, it suffices to expand the
current-density relation up to second order \cite{Spoh14}. Possible
logarithmic corrections, which arise from higher orders if $j''=0$ at
some density, are neglected \cite{BT_vanB12,Delf07}. Thus we arrive at
the nonlinear fluctuating hydrodynamics (NLFH) equation
\begin{eqnarray}
\partial_t \mathtt{u}(x,t)&=&\partial_x \left( -j^\prime\left(\rho\right)
                     \mathtt{u}(x,t)
                     -\frac{1}{2}j^{\prime\prime}(\rho)(\mathtt{u}(x,t))^2
                     \right.
                     \nonumber\\
                 &&\left. \phantom{\frac{1}{2}\qquad}
                    + \mathcal{D}\partial_x \mathtt{u}(x,t)
                    + \mathcal{B} \xi(x,t) \right).
\label{eq:NLFH_equation}
\end{eqnarray}
The noise magnitude $\mathcal{B}$ and the diffusion coefficient
$\mathcal{D}$ are related by the fluctuation-dissipation theorem
\begin{equation}
\mathcal{B}^2= 2\kappa \mathcal{D},
\label{eq:FLucDissTheo}
\end{equation}
where
\begin{equation}
\kappa=\intop \left<\mathtt{u}(0,t)\mathtt{u}(x,t)\right> \text{d}x
\end{equation}
is independent of time due to the global particle conservation. This
quantity contains information about the system's space correlations
and is a nonequilibrium analogue of the thermodynamic compressibility.

\begin{figure}
	\center{\includegraphics[width=0.45\textwidth]{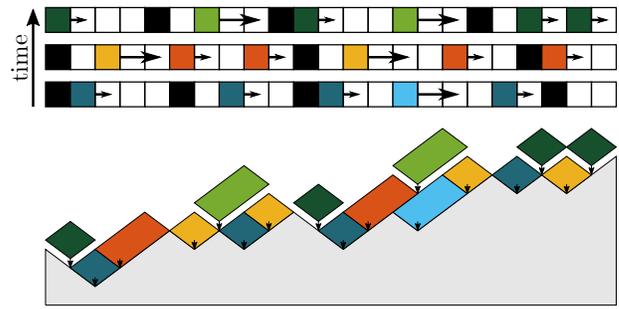}}
	\caption{ Mapping the NaSch dynamics ($v_\text{max}=2$) to 
          surface growth. Shown is a NaSch configuration evolving in
          time. The color-coded particles will hop at the next time
          step. Mapping a particle to an down-slope
          ($\blacksquare\rightarrow\diagdown$) and a hole to an
          up-slope ($\square\rightarrow\diagup$) one obtains, for each
          particle/hole configuration, a height profile. If a particle
          hops to the right a diamond is added to the surface between
          the initial and final position of the particle.}
	\label{fig:TASEP_growth_map}
\end{figure}

Notice that performing a Galilean transformation
$x\rightarrow x-v_\mathrm{col} t$ with
$v_\mathrm{col}\equiv j^\prime(\rho)$ removes the drift term
$j^\prime\left(\rho\right) \mathtt{u}(x,t)$ from the NLFH equation
(\ref{eq:NLFH_equation}) which then by writing
$\partial_xh(x,t)=-\mathtt{u}(x,t)$ turns into the originally proposed KPZ
equation \cite{KPZ}
\begin{equation}
  \partial_t h=\nu\partial_x^2 h + \frac{\lambda}{2}\left(
    \partial_x h\right)^2+\sqrt{D}\xi
\label{eq:KPZoriginal}
\end{equation}
for the surface height $h(x,t)$ with parameters
\begin{equation}
\nu=\mathcal{D},~~\lambda=-
j^{\prime\prime}(\rho)~~\text{and}~~\sqrt{D}=\mathcal{B}.
\label{eq:KPZ_NLFH_identities}
\end{equation}
For the lattice model the substitution $\partial_xh(x,t)=-\mathtt{u}(x,t)$ is
motivated by the exact mapping of the TASEP to a discrete surface
growth process
\cite{BARABASI_GROWTH,HHZ,KRUG_GROWTH,KRUG_Book_GROWTH,TASEP_GROWTH}
that is known as the single-step model. In
Fig.~\ref{fig:TASEP_growth_map} this mapping is generalized to a NaSch
scenario resulting in a growing surface with diamonds of different
size.

The universal large-scale properties of the KPZ equation are by now
well-understood, see \cite{Halp15} for a recent review.  The dynamical
exponent that relates the scaling of space and time variables as
$x\sim t^{1/z}$ takes the value $z=3/2$, as opposed to $z=1$ of the
deterministic Eulerian scaling or $z=2$ for normal diffusion.  Two
prominent exact analytic results displaying the space-time symmetry
with dynamical exponent $z=3/2$ are the asymptotic limit of the
dynamical structure function
\begin{eqnarray}
S(x,t)&=&\left<\mathtt{u}(x,t)\mathtt{u}(0,0)\right>\nonumber\\
      &\simeq& \kappa (E t)^{-\frac{1}{z}} f_\mathrm{PS}\left(
               (E t)^{-\frac{1}{z}}(x-v_\mathrm{col}t)\right)
        \label{eq:struc_fct_asymp_scaling}
\end{eqnarray}
with the Pr\"ahofer-Spohn scaling function $f_\mathrm{PS}$ \cite{Prae04} 
and the distribution
\begin{eqnarray}
  \mathcal{P}(J,t)\simeq (\Gamma t)^{-\frac{1}{2z}}
  F_\mathrm{BR}\left( -J\cdot (\Gamma t)^{-\frac{1}{2z}}\right)
\label{eq:Baik_Rains_Current_Dist}
\end{eqnarray}
of centered time-integrated currents
\begin{eqnarray}
\mathtt{J}_t&=&\intop_0^t
                [\mathtt{j}(0,s)-j(\rho)]\mathrm{d}s-\intop_0^{v_\text{col}t}
                \mathtt{u}(x,0)\mathrm{d}x
\end{eqnarray}
with the Baik-Rains scaling function $F_\mathrm{BR}(x) = 2F_0(2x)$ and
$F_0(\cdot)$ defined in \cite{BR}.  The scaling parameters are
\begin{eqnarray}
\label{eq:structure_fct_scaling_factor}
E&=&|j^{\prime\prime}|\sqrt{2\kappa}\\
\Gamma&=&|j^{\prime\prime}|4\kappa^2.
\end{eqnarray}
The exactly known scaling functions $f_\mathrm{PS}$ and
$F_\mathrm{BR}$ are given by solutions of of certain Painlev\'e II
transcendent equations \cite{PJF03}, and cannot be expressed in closed
form but are tabulated with high precision \cite{Prae04_DATA}.

In order to check whether the NaSch model is in the KPZ universality
class we first calculate the hydrodynamic quantities $\kappa$, $j$,
$v_\mathrm{col}$ and $j^{\prime\prime}$ exactly for $v_\text{max}=1$
and from Monte-Carlo simulations for $v_\text{max}>1$. This allows us
to fix the analytic predictions for the dynamical structure function and
time-integrated current distribution for comparison with numerical
data. We stress that the quantities $\kappa$, $v_\text{col}$ and
$j^{\prime\prime}(\rho)$ are {\it purely stationary} quantities that
do not require any knowledge about space-time symmetry. Thus, a
comparison of simulation results for the dynamical structure function and
current statistics with the analytical predictions
(\ref{eq:struc_fct_asymp_scaling}) and
(\ref{eq:Baik_Rains_Current_Dist}) serves as a reliable check whether
the NaSch model truly belongs the KPZ universality class.

As the discussion so far treats the dynamics as continuous in space
and time we need to define for the NaSch model a discrete version of
the hydrodynamic quantities $\rho$, $j(\rho)$, $\kappa$, the structure
function and time-integrated currents. To this end a configuration
$\mathcal{C}(t)=\{n_{x,t},v_{x,t}\}$ of the NaSch model at the end of
an update cycle $t$ is expressed by a pair of occupation numbers
$n_{x,t}\in\{0,1\}$ and its associated velocities
$v_{x,t}\in\{0,\ldots,n_{x,t}\cdot v_\text{max}\}$ at site $x$. We
limit our simulations to periodic systems of length $L$ with fixed
particle density $\rho=\frac{1}{L}\sum_{x=1}^L n_{x,t}$.  The
current-density relation is calculated as
$j(\rho)=\rho\left< v \right>$ where $\left< v \right>$ is the
stationary average velocity of the cars. For $v_\text{max}>1$ the
stationary state is unknown and the compressibility is calculated from
space correlations as
\begin{eqnarray}
\kappa&=&\sum_{x=-K}^{K} \left( \left< n_{0,t}n_{x,t} \right>-\rho^2 \right)\\
&=&\sum_{x=-K}^{K}S(x,0)
\end{eqnarray}
where the cutoff $K\ll L/2$ excludes exponentially decaying space
correlation which can be neglected within statistical accuracy.  With
the hydrodynamic quantities at hand the dynamical structure function
\begin{equation}
\label{eq:SingleLaneStrucFCT}
S(x,t)=\left< n_{x,t} n_{0,0} \right>-\rho^2
\end{equation}
can be measured and compared with the scaling form
(\ref{eq:struc_fct_asymp_scaling}).  To define a discrete version of
the centered time-integrated current we have to introduce a discrete
version of the instantaneous current
\begin{equation}
  \mathtt{j}_{x,t}=\sum_{x^\prime=x+1}^{x+v_\mathrm{max}}
  \sum_{p=x^\prime-v_{x^\prime,t}}^{x^\prime-1}\delta_{x,p} 
\end{equation}
indicating if a particle passes between sites $x$ and $x+1$ during the
update from $t-1$ to $t$.  Finally, the discrete centered time-integrated
current satisfying Eq.~(\ref{eq:Baik_Rains_Current_Dist}) is given as
\begin{equation}
\label{eq:time_integrated_current}
J_{x,t}=\sum_{s=1}^{t}[\mathtt{j}_{x,s}-j(\rho)]-\sum_{x^\prime=0}^{\lfloor
  j^\prime(\rho)t\rfloor}[n_{x+x^\prime,0}-\rho].
\end{equation}
With these quantities we are in a position to probe in detail the
dynamical universality class of the NaSch model.

\section{NaSch Model with $v_{\text{max}}=1$}
\label{sec::TASEPResults}

For $v_{\text{max}}=1$ the NaSch model corresponds to the TASEP with
parallel dynamics. In this case, the stationary state is exactly known
allowing to determine all nonuniversal scaling factors
exactly. Therefore, this special case serves as a benchmark to show
the convergence towards the asymptotic limit and allows to identify
early time contributions. The latter will play a key role to point out
that early time contribution might persist longer than expected. This
leads to essential insights into finite-time and -size effects for
simulations with unknown steady state ($v_{\max}>1$).

\begin{figure}[h]
	\centerline{\includegraphics[width=0.45\textwidth]{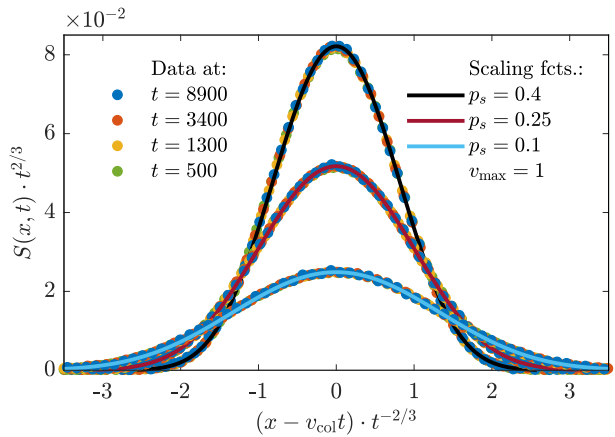}}
	\caption{Scaling plot with dynamical exponent $z=3/2$ for the
          measured dynamical structure function of a TASEP system with parallel
          update rule and various hopping probabilities $p=1-p_s$. The
          comparison to the asymptotic scaling form
          (\ref{eq:struc_fct_asymp_scaling}) with analytically
          obtained scaling factor
          (Eqs.~(\ref{eq:structure_fct_scaling_factor}), (\ref{eq:NaSch_vmax_1_current_density_relation})
          and (\ref{eq:NaSch_vmax_1_compressibility})) shows a
          remarkably agreement although it was obtained for a model
          continuous in time and space. The probability sequence $p_s$
          in the legend matches the scaling function sequence from top
          to bottom. The TASEP system parameters are $\rho=1/2$,
          $L=10^7$ whereas the initial configuration was drawn from to
          the exactly known stationary distribution
          (\ref{eq:2,1-Cluster-Ansatz})-(\ref{eq:NaSch_vmax_1_state_distribution}).
          The Monte Carlo parameters are $P=100$, $M=100$, $\tau=100$
          resulting in statistical errors of the order of the symbol size. For
          better visibility not all data points are shown.  }
	\label{fig:Struc_Fct_Collapse_vmax_1_rho_500}
\end{figure}

Interpreting $p=1-p_s$ as the probability that a vehicle will move, we
have full equivalence to the TASEP with parallel update rule.  Hereby
the random-sequential update is included as a limiting case when
taking in the limit of $p\rightarrow 0$ (time properly rescaled),
while in the limit of $p\rightarrow 1$ the dynamics become
deterministic.  The exact stationary probability distribution
$\bar{P}(\mathbf{n})$ to observe a configuration $\mathbf{n}$
factorizes into a two cluster form
\begin{equation}
\label{eq:2,1-Cluster-Ansatz}
\bar{P}\left(\mathbf{n}\right)=\prod_{x=-\infty}^{\infty}P\left({n}_{x},{n}_{x+1}\right),
\end{equation}
where $n_k\in{0,1}$ is the occupation number at site $k$.
Using the Kolomogorov consistency relations
\begin{eqnarray}
P(0,0)&=&P(0)-P(1,0)\\
P(1,1)&=&P(1)-P(1,0)\\
P(0,1)&=&P(1,0)
\end{eqnarray}
with $P(1)=\rho$ and $P(0)=1-\rho$ one consequently has to solve the
master equation for $P(1,0)$. Expressed in terms of $P(1,0)$ the
stationary master equation reduces to a quadratic form and yields
\begin{equation}
\label{eq:NaSch_vmax_1_state_distribution}
P(1,0)=\frac{1}{2(1-p_s)}\left[ 1- \sqrt{1-4(1-p_s)\rho(1-\rho)} \right].
\end{equation}
With the stationary distribution at hand, one calculates the 
current-density relation and its compressibility as
\begin{eqnarray}
\label{eq:NaSch_vmax_1_current_density_relation}
j(\rho)&=&\frac{1}{2}\left[ 1- \sqrt{1-4(1-p_s)\rho(1-\rho)} \right]\\
\label{eq:NaSch_vmax_1_compressibility}
\kappa &=& \rho(1-\rho)\sqrt{1-4(1-p_s)\rho(1-\rho)}
\end{eqnarray}
fixing the nonuniversal scaling parameter. Knowing these hydrodynamic
quantities exactly we are in the position to compare simulation
results to the exact asymptotic predictions derived in
Sec.~\ref{sec:NLFH} without any free parameter left.
Fig.~\ref{fig:Struc_Fct_Collapse_vmax_1_rho_500} shows a scaling plot
with dynamical exponent $z=3/2$ of simulation data obtained for
$\rho=1/2$ and various $p_s$. Additionally, in
\cite{Meersoon_Schmidt_17} the parallel update TASEP has been shown to
exhibit the Baik-Rains distribution
(\ref{eq:Baik_Rains_Current_Dist}) for current fluctuations.
Remarkably, the data for the dynamical structure function and current
distribution matches the predicted scaling form perfectly, although it is obtained for a model continuous in time and space.\\

\section{NaSch Model with $v_{\text{max}}>1$}
\label{sec::NaSchResults}
As mentioned before the stationary state is unknown for
$v_\text{max}>1$.  Thus, the system has to be relaxed before one
starts recording observables using the Metropolis sampler.
Especially, it has been shown that the KPZ statistics are sensitive to
initial conditions and might reveal different scaling functions
\cite{Meersoon_Schmidt_17,Ferrari_Spohn}.  For a stochastic model the
relaxation time $T_\mathrm{relax}$ can often be defined through the
spectral gap $G_\mathrm{S}$ of the time evolution operator
\cite{SCNBook} which depends on the system size as
$T_\mathrm{relax}\sim G_\mathrm{S} \sim L^{z}$, where $z$ is the
dynamic exponent. For the ASEP with periodic and open boundaries the
spectral gap was calculated exactly using Bethe ansatz methods
\cite{Dhar,Gwa92,Kim95,Gier_Essler}.

Since the spectral gap is in general hard to calculate, one might use
the dynamical structure function to define an equivalent relaxation
time. The dynamical structure function carries the information about
the slow relaxation mode and displays the evolution of a
perturbation/fluctuation trough the system. The amplitude of the
dynamical structure function will decay exponentially instead of
$t^{-1/z}$ after the dynamical structure function has been spread over
the whole system \cite{Prolhac_16}. The width $\sigma$ of the
dynamical structure function scales with time as $\sigma\sim
t^{1/z}$. In case of the KPZ universality class we define the width as
$\sigma(t)\equiv (\sqrt{2\kappa}|j^{\prime\prime}|t)^{2/3}$ covering
$\intop_{-0.5}^{0.5}f_{\text{PS}}(x)\text{d}x=50.057\ldots\%$ of the
Pr\"ahofer Spohn KPZ scaling~function, whereas one has
$f_\text{PS}(0)=2f_\text{PS}(\pm0.88046626\ldots)$.  Thus, a lower
boundary for a proper relaxation time is given when the structure
function width $\sigma$ covers the whole system. Solving
$L\lesssim(\sqrt{2\kappa}|j^{\prime\prime}|T_\mathrm{relax})^{2/3}$ we
derive the relaxation time as
\begin{eqnarray}
\label{eq:relax_time}
T_\mathrm{relax}&\gtrsim
  &\frac{L^{3/2}}{\sqrt{2\kappa}|j^{\prime\prime}|}
    \,.
\end{eqnarray}
It turns out that the system's relaxation is the major computation
bottleneck. A propper bound for a minimum required relaxation time
allows a significant reduction of computation cost, and ensures
relaxation artifacts to be absent. Because, the derived relaxation
bound Eq.~(\ref{eq:relax_time}) applies for systems near the
stationary state, we introduce a two level relaxation. Thus, we first
initialize the system with equally spaced vehicles, velocity
$v=v_\text{max}$, pre-relax the state according to
Eq.~(\ref{eq:relax_time}) and store it in memory. In this way, the
chosen initial condition prevents the system of being stuck in a jam
which may have a long life-time \cite{Nagel_Jam_93}.  Second, to
generate a new independent state, the relaxed state is loaded and
again independently propagated according to Eq.~(\ref{eq:relax_time}).
However, there are various ways to initialize the system which may
have advantages or disadvantages, depending on the observed
quantities.  We have tested our data for independence on the initial
state by choosing different initial conditions and comparing the
observables.  Only for $T_\mathrm{relax}\gtrsim L^{3/2}/E$ these
differences disappear.

\begin{figure}[h]
\centerline{\includegraphics[width=0.45\textwidth]{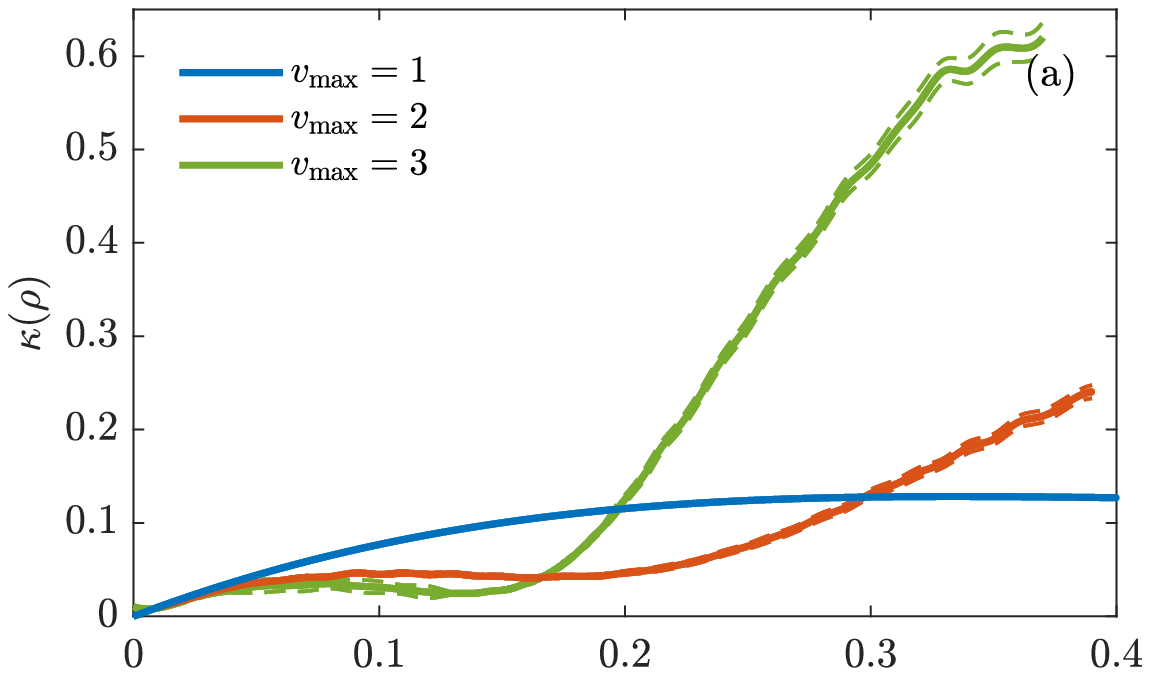}}
\centerline{\includegraphics[width=0.45\textwidth]{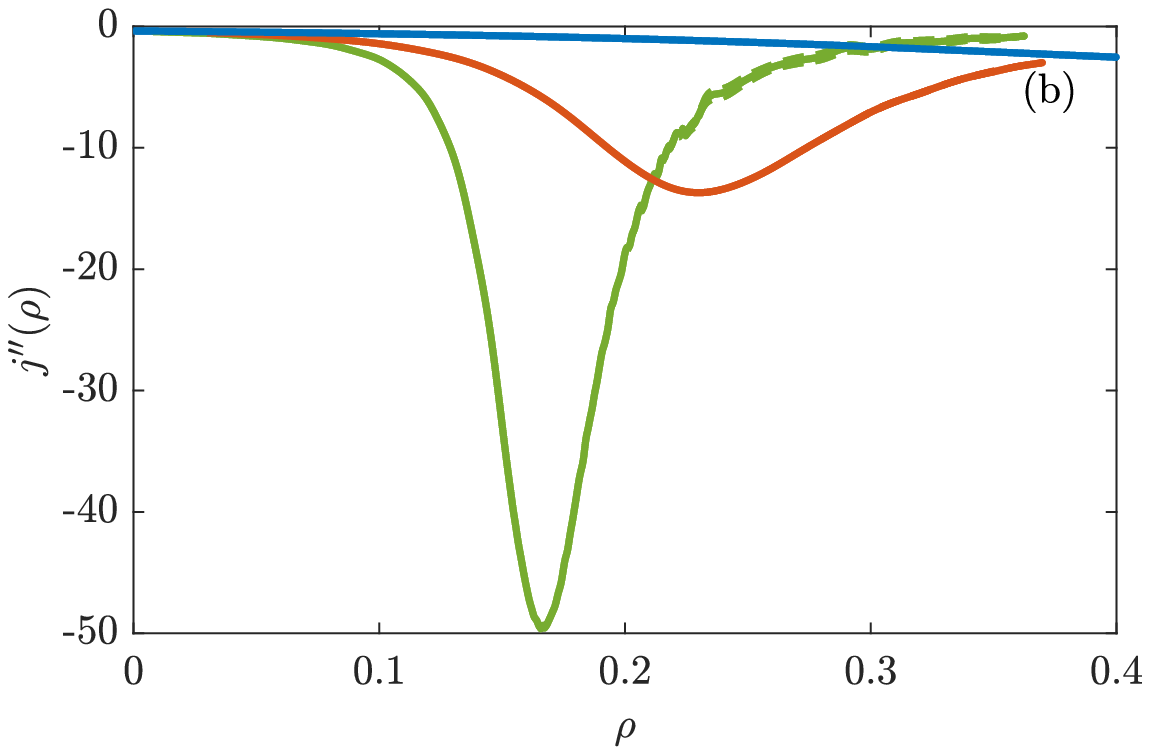}}
\caption{(a) Compressibility plot for the NaSch model with $p_s=0.25$
  and $v_\text{max}\in\{1,2,3\}$. The compressibility-density diagram
  for the NaSch model exhibit for $v_{\text{max}}>1$ two maxima and
  one minimum ($\kappa(1)=0$). The formation of spontaneous traffic
  jams cause an enhanced compressibility.\\
  (b) Second derivative of
  current-density relation for $p_s=0.25$ and various
  $v_{\text{max}}$. The second derivative was calculated from current
  data using finite-difference formulas with accuracy
  $\mathcal{O}(h^8)$, where $h$ is the grid spacing.
  For $v_\text{max}>1$ the data is recorded in systems of size
  $L=200.000$ using ergodic measurements ($\tau=1$) after relaxing the
  system according to Eq.~(\ref{eq:relax_time}) and averaging over
  independent realizations. Dashed lines indicate the 99\% confidence
  bound.  }
\label{fig:Compressibility_current_derivative}
\end{figure}

Fig.~\ref{fig:Compressibility_current_derivative} shows the
compressibility and the second derivative of the current 
as function of the density for different values of
$v_\text{max}$. The behaviour of the compressibility for
$v_{\text{max}}>1$ differs clearly from that for $v_{\text{max}}=1$.
In the latter case, $\kappa(\rho)$ increases monotonically with
increasing density $\rho$ whereas for $v_{\text{max}}>1$ two local
extrema exist in the interval $0 < \rho < 1$.
The compressibility is strongly enhanced at higher densities,
reflecting the formation of spontaneous traffic jams.

\begin{figure}[h]
\centerline{\includegraphics[width=0.45\textwidth]{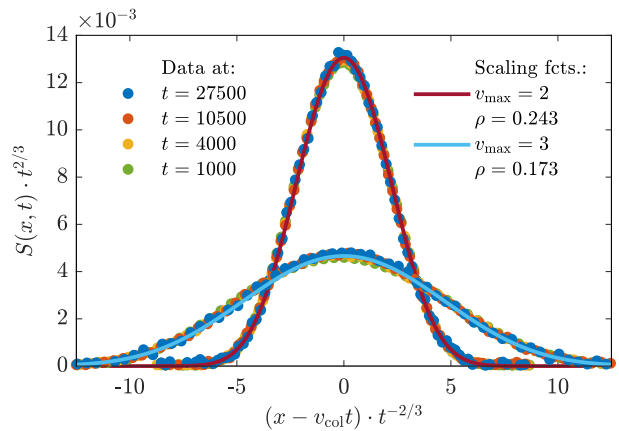}}
\caption{The data collapse for the NaSch ($p_s=0.25$)
          dynamical structure function shows a nice aggrement with the
          asymptotic scaling function
          (\ref{eq:struc_fct_asymp_scaling}). Choosing
          $\rho\approx\rho^\star$, the dynamical structure function does not
          show a skew at early times, compare to
          Fig.~\ref{fig:Time_Collapse_Asymmetry}.  The Monte-Carlo
          parameters for the dynamical structure function are $L=200.000$,
          $P=300$, $\tau=500$ and $M=50.000$. The hydrodynamic
          quantities $\kappa$, $v_\mathrm{col}$,
          $\partial_{\rho}^{2}j$ are measured using separate and
          independent Monte-Carlo simulations (see
          Fig.~\ref{fig:Compressibility_current_derivative} for
          details). The parameters are for $v_\text{max}=2$ (upper)
          $\kappa=0.07\pm0.001$, $v_\mathrm{col}=0.6308\pm0.0003$,
          $\partial_{\rho}^{2}j=-13.22 \pm0.04$ and for
          $v_\text{max}=3$ (lower) $\kappa=0.0524\pm0.0008$,
          $v_\mathrm{col}=1.0513\pm0.0002$,
          $\partial_{\rho}^{2}j=-46.6\pm0.1$.  For better visibility not all
          data points are shown. Statistical errors are of the order of
          the symbol size.  }
	\label{fig:Structure_function_collapse_vmax_2_3_sym}
\end{figure}

\begin{figure}[h]
	\centerline{\includegraphics[width=0.45\textwidth]{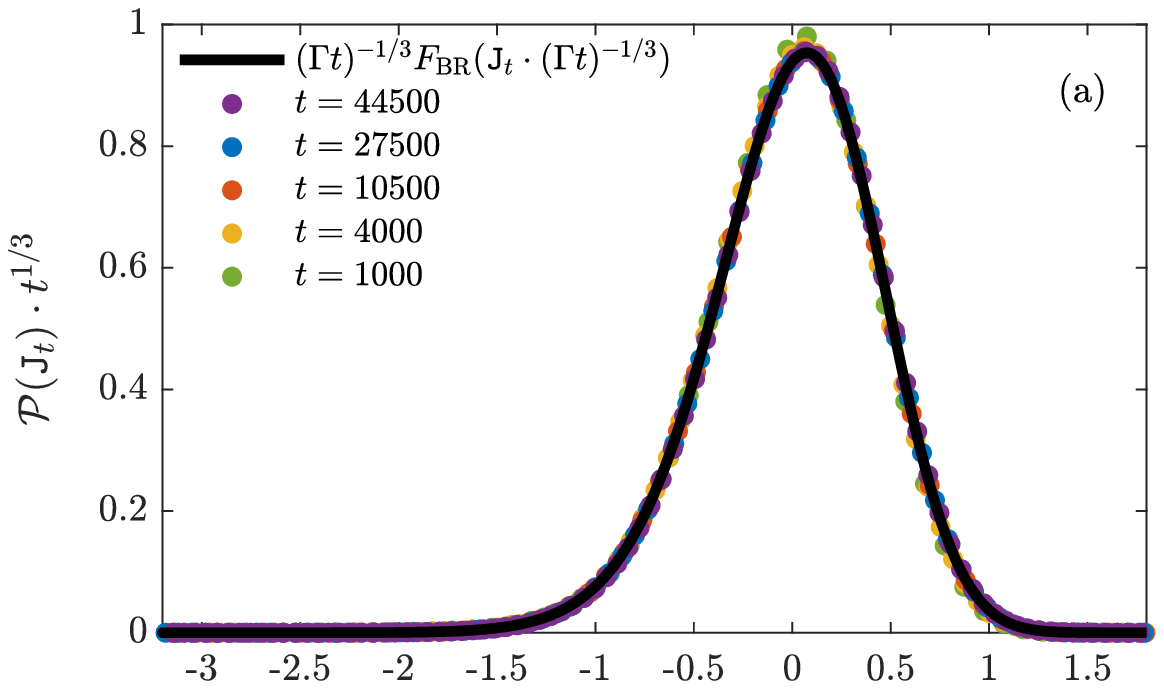}}
	\centerline{\includegraphics[width=0.45\textwidth]{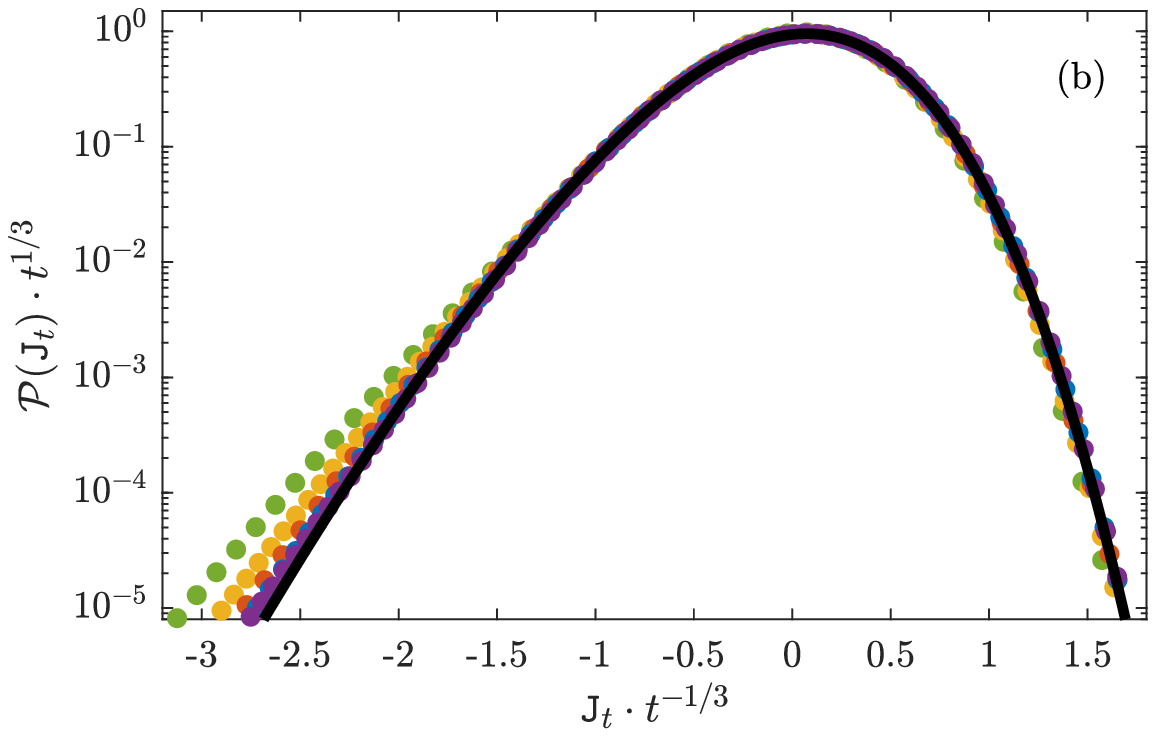}}
        \caption{The time collapse of the distribution of time
            integrated currents (\ref{eq:time_integrated_current}) for
            the NaSch model ($v_\text{max}=3$, $p_s=0.25$, $\rho=0.173$)
            shows a nice agreement with the asymptotic Baik-Rains
            distribution (\ref{eq:Baik_Rains_Current_Dist}). The
            Monte-Carlo parameters for recorded current distributions
            are $L=200.000$, $P=500$, $\tau=1$ and $M=6.000.000$. The
            hydrodynamic quantities are independently measured as
            $\kappa=0.0524\pm0.0008$, $j=0.43214\pm2\cdot10^{-6}$,
            $v_\mathrm{col}=1.0513\pm0.0002$,
            $\partial_{\rho}^{2}j=-46.6\pm0.1$ (for details see
            Fig.~\ref{fig:Compressibility_current_derivative}).  For
            better visibility, not all data points are shown. Statistical
            errors are of the order of the symbol size.  }
	\label{fig:Current_distribution_vmax_3_sym_p_025}
\end{figure}

The data for the dynamical structure functions
(Fig.~\ref{fig:Structure_function_collapse_vmax_2_3_sym}) for
$v_{\text{max}}>1$ and $\rho\approx\rho^\star$, where
$\rho^{\star}:=\text{argmax}_{\rho\in(0,1)}E$ is the density for which
the scaling parameter $E$ becomes maximal, collapse well and show a
very good agreement with the asymptotic scaling function
(\ref{eq:struc_fct_asymp_scaling}).  The time collapse of the
distribution for the time-integrated currents
(\ref{eq:time_integrated_current}) in
Fig.~\ref{fig:Current_distribution_vmax_3_sym_p_025} shows a nice
agreement with the asymptotic Baik-Rains distribution
(\ref{eq:Baik_Rains_Current_Dist}).

\section{Early time dynamical structure function \label{sec:Early_Time}}

NLFH has produced only asymptotic results so far. A full space-time
solution of Eq.~(\ref{eq:NLFH_equation}) would allow for a better
comparison with simulation data and therefore a better identification of
corrections which may arise from higher order corrections to
Eq.~(\ref{eq:NLFH_equation}).  In this section we will take a closer
look on simulation data for the early time dynamical structure function, showing
a density-dependent asymmetry that vanishes with time. Non-asymptotic
effects might have a strong impact on the identification of universal
behavior and may lead to inconclusive results. Therefore, a
qualitative understanding of the early time dynamical structure function
asymmetry is crucial for the interpretation of simulation data.

In order to easily compare data for different models and parameters,
the dynamical structure functions are rescaled to its scaling function as
\begin{eqnarray}
\kappa^{-1}(E t)^{2/3}S\left((E t)^{2/3} x + j^{\prime}(\rho)
  t,t\right) &\simeq & f_{\text{PS}}(x).
\label{eq:PS_SCALING_AND_DATA}
\end{eqnarray}

\begin{figure}[h]
\centerline{\includegraphics[width=0.45\textwidth]{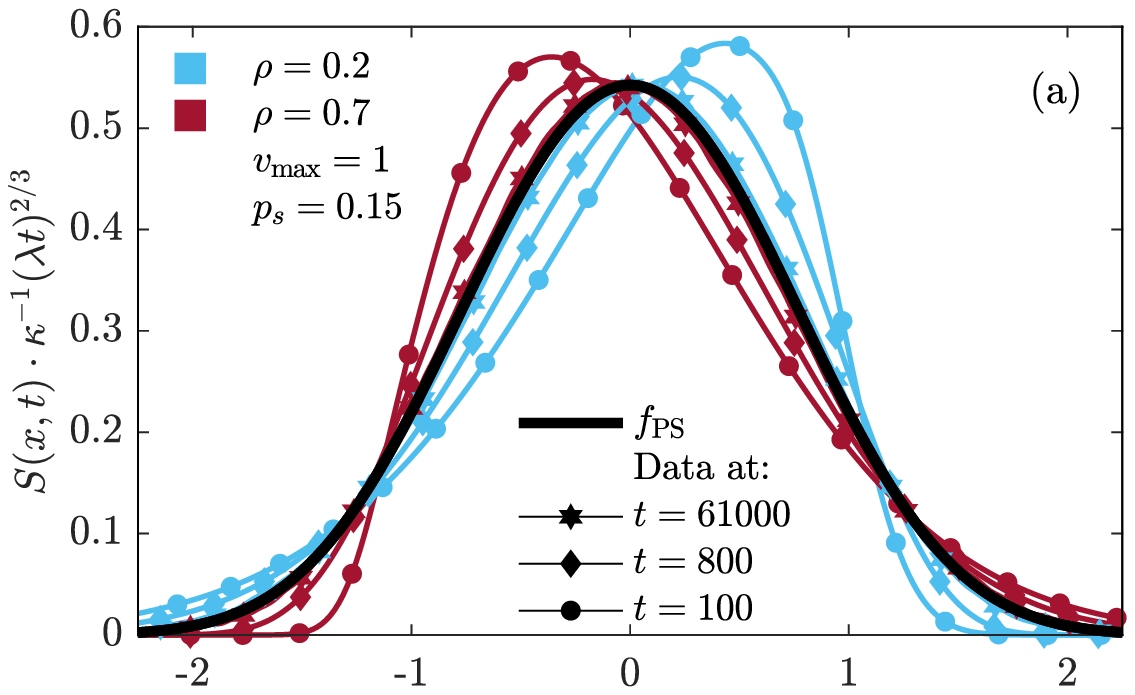}}
\centerline{\includegraphics[width=0.45\textwidth]{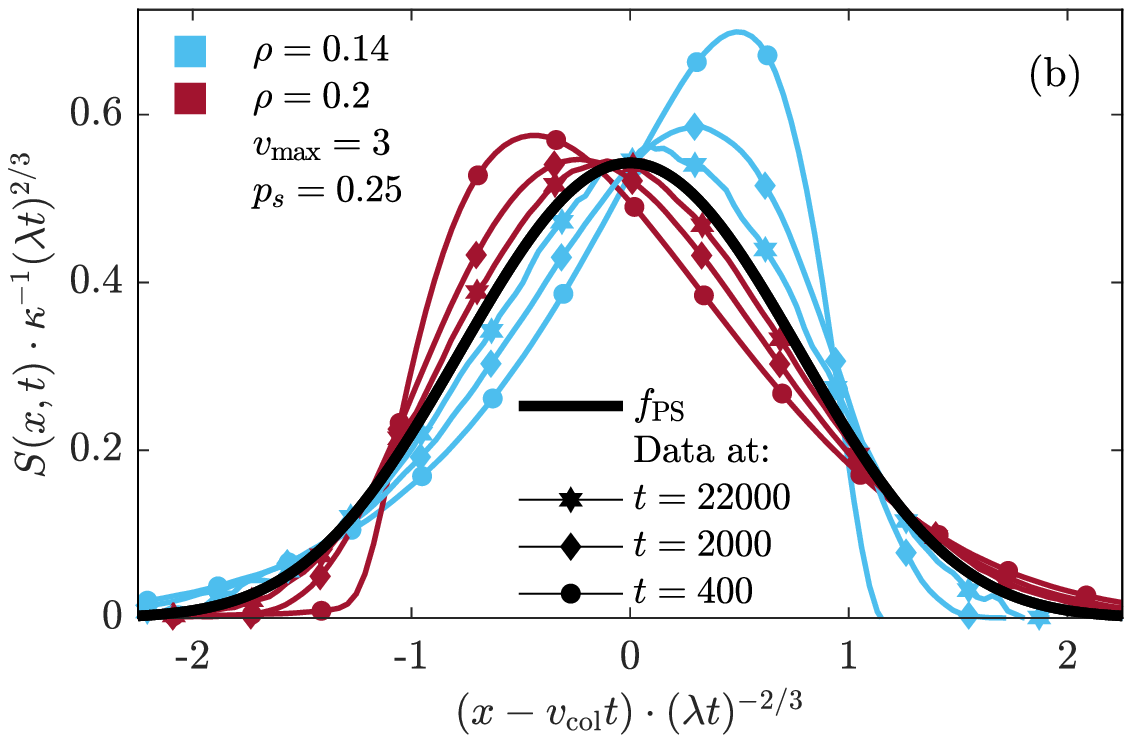}}
\caption{Structure-function scaling plots compared to the asymptotic
  KPZ solution (\ref{eq:PS_SCALING_AND_DATA}).  The measured
  dynamical structure function shows for the TASEP (a)
  and the NaSch model (b) a skew present at early times for densities $\rho\not=\rho^\star$ and slowly converges to the expected scaling behaviour.\\
  (a) TASEP (i.e. NaSch model with $v_{\text{max}}=1$) for $p_s=0.15$ and
  densities $\rho=0.2$ (left), $\rho=0.7$ (right).
  The Monte-Carlo parameters for the simulations are $L=10^7$, $P=300$, $\tau=100$ and $M=500$.\\
  (b) NaSch model with $v_{\text{max}}=3$, $p_s=1/4$ and densities
  $\rho=0.14$ (left), $\rho=0.2$ (right).  The Monte-Carlo parameters
  are $L=200.000$, $P=300$, $\tau=400$ and $M=50.000$.  The
  hydrodynamic quantities $\kappa$, $v_\mathrm{col}, $ are measured
  using separate and independent Monte-Carlo simulations (see
  Fig.~\ref{fig:Compressibility_current_derivative}).  The parameters
  are for $\rho_1=0.14$: $\kappa_1=0.0249\pm0.0005$,
  $v_{\mathrm{col},1}=2.3376\pm0.0001$,
  $\partial_{\rho}^{2}j=-19.17 \pm0.04$
  and for $\rho_2=0.2$: $\kappa_2=0.125\pm0.002$, $v_{\mathrm{col},2}=0.1878\pm0.0003$, $\partial_{\rho}^{2}j=-18.7\pm0.1$.
  For better visibility not all data points are shown. Thin lines are guides
  for the eye and statistical errors are of the order of the symbol
  size.}
	\label{fig:Time_Collapse_Asymmetry}
\end{figure}

Due to the particle-hole symmetry of the TASEP the measured dynamical
structure function is symmetric for $\rho=1/2$ and matches the
symmetry prediction of the asymptotic solution
Eq.~(\ref{eq:struc_fct_asymp_scaling}).  However, for densities
$\rho\not=1/2$, the early time dynamical structure function shows an
asymmetry which vanishes with increasing time.  As shown in
Fig.~\ref{fig:Time_Collapse_Asymmetry} the asymmetry is present both
for $v_\text{max}=1$ (TASEP) and $v_\text{max}>1$. Thus the asymmetry
is not a special feature of the NaSch model where particle velocities
might be interpreted as an internal degree of freedom.  Note that
fitting the dynamical exponent from the maximum of the dynamical
structure function in a non-asymptotic regime will lead to
density-dependent and therefore inconclusive results as observed in
\cite{CsanyiK95,SasvariK97}.

\begin{figure}[h]
\centerline{\includegraphics[width=0.45\textwidth]{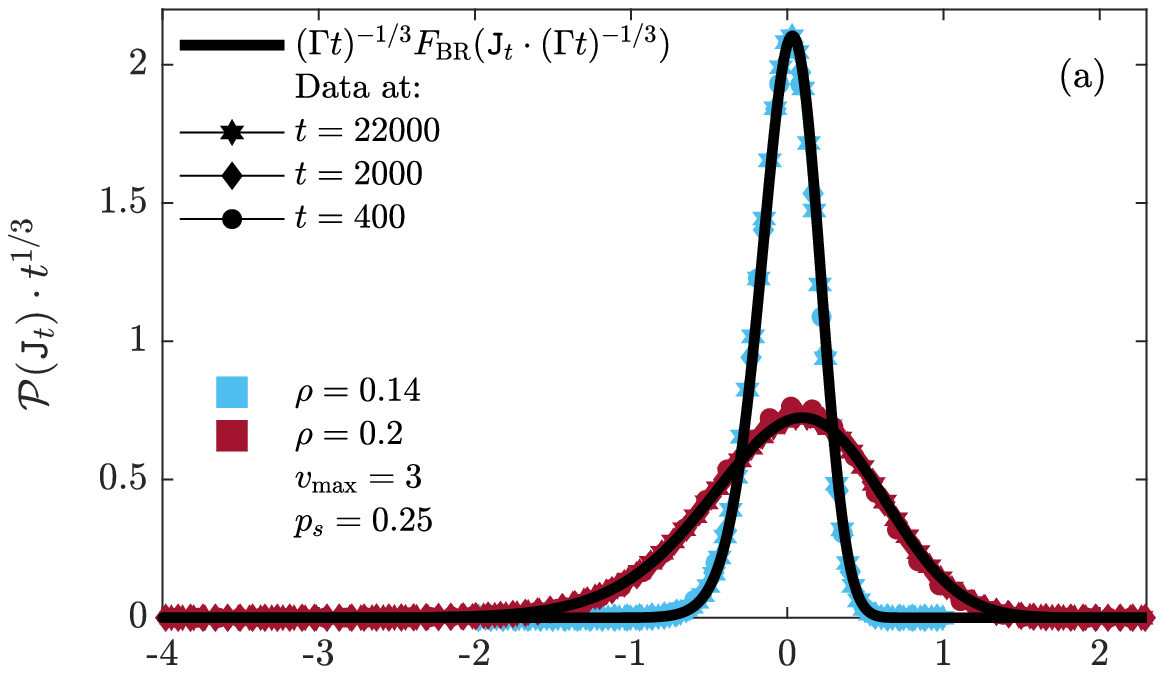}}
\centerline{\includegraphics[width=0.45\textwidth]{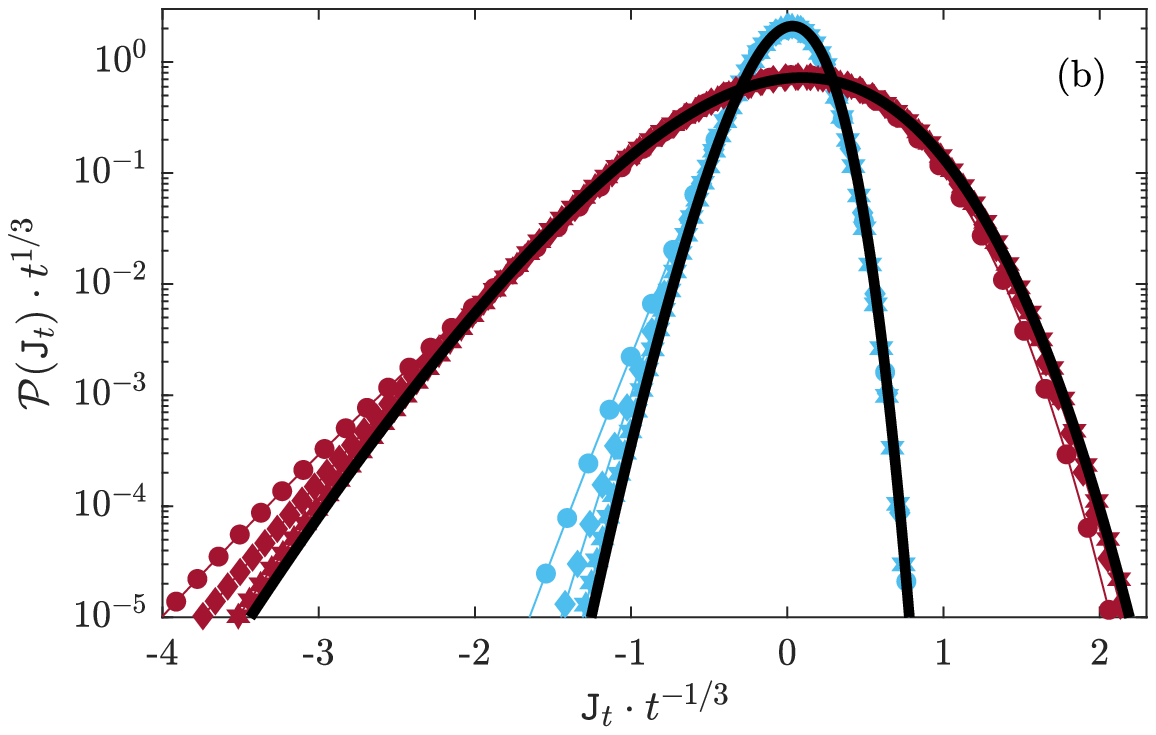}}
\caption{Taken the NaSch models from
  Fig.~\ref{fig:Time_Collapse_Asymmetry}b) showing a skewed
  dynamical structure function, the time collapse for time-integrated currents
  distributions (\ref{eq:time_integrated_current}) shows a nice
  agreement with the asymptotic Baik-Rains distribution
  (\ref{eq:Baik_Rains_Current_Dist}) and is more stable against
  finite time corrections.  The Monte-Carlo parameters for the current
  distribution are $L=200.000$, $\tau=1$, $M=20.000.000$ and
  $P=500$. For better visibility not all data points are shown. Thin lines
  are guides for the eye and statistical errors are of the order of the symbol
  size.}
	\label{fig:Time_Collapse_Current_Asymmetry}
\end{figure}

The skew of the dynamical structure function that we observe at early
times disappears for $\rho \approx \rho^\star$. It is negative for
$\rho<\rho^\star$ and positive for $\rho>\rho^\star$
(Fig.~\ref{fig:Time_Collapse_Asymmetry}) and increases with
$|\partial_\rho E|$. This indicates the role of cubic corrections to
Eq.~(\ref{eq:NLFH_equation}) for the full time solution of the
dynamical structure function. On the other hand, the distribution of
the time-integrated current does not show indications for higher order
corrections (Fig.~\ref{fig:Time_Collapse_Current_Asymmetry}).

\section{Two-Lane NaSch Model with Dynamical Lane Changes}

In order to further understand the relevance of universal behavior
for traffic-like models we will now relax the condition of single-file
motion.
We consider a one-dimensional system with two lanes and 
dynamical symmetric lane changing rules that allow overtaking on both
lanes.
For our purpose, we do not need rules that lead to a very realistic
simulation of multilane traffic \cite{Rickert,NagelWWS,SCNBook},
but represent only the basic aspects of lane changing.

Generically, a lane change decision is based on two criteria: The
\textit{incentive criterion} which tests for an improvement of the
individual traffic situation, e.g. to move forward with their desired
velocity, and the \textit{safety criterion} where each vehicle considers
a lane change based on the available backward gap in the
desired lane \cite{Rickert,NagelWWS,SCNBook}. 

It is natural to split the multi-lane-model update into two substeps: In
the first substep vehicles may change lanes and in the second
substep vehicles move forward as in the single-lane NaSch model.

\begin{figure}[h]
\centerline{\includegraphics[width=0.40\textwidth]{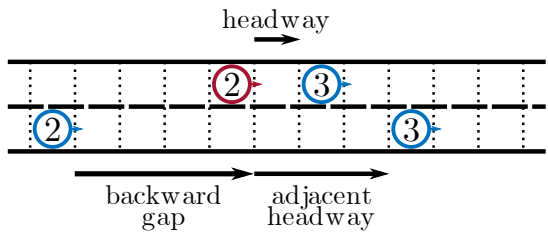}}
\caption{Schematic drawing of a two-lane Nagel-Schreckenberg model
  with dynamical lane-changing rule. The configuration shows of a
  typical lane change situation where the incentive and safety
  criterion are satisfied. The configuration is shown at the start of
  the update cycle and numbers indicate the vehicles velocities. As
  the particles marked in red can not advance with its desired speed it
  considers a lane change.}
\label{Lane_Change_Example}
\end{figure}

The investigated lane change protocol is designed as follows:
\begin{itemize}
\item \textit{Incentive criterion:} If the headway $d_n$ in front of
  the $n$-th vehicle on lane $\lambda$ $(\lambda =1,2)$ is too small
  to travel with the desired speed in the ensuing NaSch update and the
  headway $d_n^{\text{(a)}}$ in the adjacent lane is larger, the
  vehicle considers a lane change. Otherwise, it stays in its actual
  lane, i.e.
\[
d_n<\min\left(v_n+1,v_{\text{max}}\right)~\text{AND}~d_n<d_n^{\text{(a)}}
\]

\item \textit{Safety criterion:} The $n$-th vehicle got a neighboring
  vehicle on the adjacent lane which might be next to or behind to
  it. This neighboring vehicle is moving with velocity
  $v_n^\text{(a)}$ and $b_n$ measures the backward gap. The backward
  gap is equal to zero if the vehicles are next to each other. To
  avoid conflicts due to lane changes in the following NaSch update,
  the backward gaps should be sufficiently large, so that neighboring
  cars won't break due to the lane changes, i.e.
\[
b_n>\min(v_n^{\text{(a)}}+1,v_{\text{max}})
\]

\item \textit{Randomization:} If the criteria above are satisfied the
  vehicle performs a lane change with probability $p_c$
\end{itemize}
Lane changes are performed in parallel. Fig.~\ref{Lane_Change_Example}
shows a typical lane change situation.

\begin{figure}[h]
	\centerline{\includegraphics[width=0.45\textwidth]{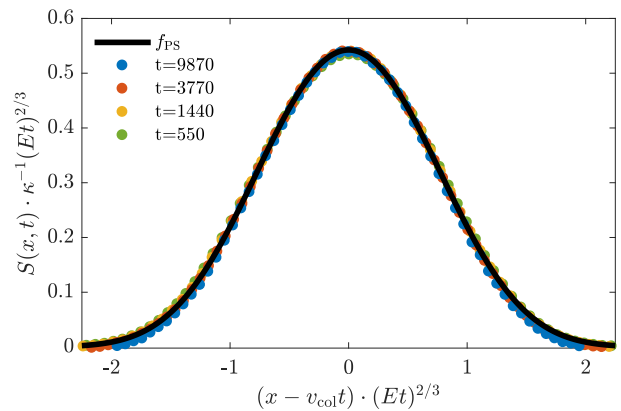}}
	\caption{Time collapse for measured two-lane NaSch structure
          function (\ref{eq:two-lane-structure-function}) shows a nice
          agreement with the asymptotic NLFH solution
          (\ref{eq:struc_fct_asymp_scaling}). For times $t\geq 80$ on-
          ($\lambda=\mu$) and cross-lane ($\lambda\not=\mu$) structure
          functions $S^{\lambda \mu}$ do not differ from each other
          within statistical accuracy. This indicates that the relaxation
          between lanes does not contribute to the long term relaxation
          behaviour.  The quantities $v_\mathrm{col}$ and
          $j^{\prime\prime}$ are calculated independently using
          finite-difference formulas and current data for different
          densities. The compressibility $\kappa$ is determined from
          the dynamical structure function at $t=0$ (space-correlations). The
          simulated model parameter are $\rho=1/4$, $p_s=1/4$,
          $p_c=1/2$ and $L=200,000$ resulting in
          $\kappa=0.0308\pm0.0002$,
          $v_{\mathrm{col}}=0.65433\pm0.00005$ and
          $j^{\prime\prime}=-14.43\pm0.01$. The Monte-Carlo parameters
          are $L=200.000$, $P=2000$, $\tau=10$ and $M=25000$. For
          better visibility not all data points are shown. Statistical
          errors are of the order of the symbol size. }
	\label{2Lane_plot4}
\end{figure}

Note that all vehicles are identical and the system only conserves the
overall vehicle density, therefore the system is expected to support a
single KPZ-mode.  Accounting for the symmetry of the model, the structure
function and its hydrodynamic quantities can be defined as
\begin{eqnarray}
S^{\lambda \mu}(x,t) &=& \left< n^{\lambda}_{x,t}n^{\mu}_{0,0}\right> - \rho^2
\end{eqnarray}
\begin{eqnarray}
\label{eq:two-lane-structure-function}
S(x,t) &=& \frac{1}{4}\sum_{\lambda , \mu =1}^{2} S^{\lambda \mu}_{k}(t)
\end{eqnarray}
\begin{eqnarray}
\kappa &=& \frac{1}{4}\sum_{\lambda , \mu =1}^{2} \sum_{x=-K}^K S^{\lambda \mu}(x,t)
\end{eqnarray}
\begin{eqnarray}
\rho &=&\frac{1}{2L}\sum_{x=1}^L\left( n^{1}_{x,t} +n^{2}_{x,t}\right)\\
j(\rho) &=& \frac{1}{2}\sum_{\mu =1}^{2}\rho\left< v^{\mu}_{x,t} \right>.
\end{eqnarray}
In Fig.~\ref{2Lane_plot4} a nice agreement between Monte-Carlo
simulations and predicted asymptotic KPZ scaling behaviour
(\ref{eq:struc_fct_asymp_scaling}) is shown. In order to reach the
asymptotic regime within computation limits a vehicle density
$\rho \approx \rho^\star$ was used to avoid an early time
skew of the dynamical structure function.

The results of this section once again show the robustness of
the KPZ universality class. To observe KPZ behavior, single-file motion
is not a necessary condition.

\section{Discussion}

We have provided strong numerical evidence that the NaSch model of
traffic flow belongs to the KPZ universality class for all choices of
the parameters $v_{\text{max}}$ and $p$. Previously, this was only
known for the special limit $v_{\text{max}}=1$ and random-sequential
dynamics where the model corresponds to the TASEP. Previous studies
\cite{CsanyiK95,SasvariK97}
were unable to determine the universality class conclusively because
of the strong finite-size and finite-time effects (see
Fig.~\ref{fig:Time_Collapse_Asymmetry}).

The results presented here provide deeper insights both in the
universality of driven diffusive systems and the dynamics of traffic
flow. They indicate that neither the updating procedure (parallel
update for NaSch vs. random-sequential update for TASEP) nor the
internal degree of freedom (i.e. the velocity which is introduced for
$v_{\text{max}}>1$) affect the universality. Furthermore, we have shown
by considering a multi-lane version of the NaSch model that the
universality class is also not changed by deviations from strict
single-file motion, i.e. allowing for changes in the particle
ordering. 

To test for KPZ universality the dynamical structure function and the
distribution of time-integrated currents were recorded for various
times. Both observations show a nice agreement with the analytical
predictions.
The dynamical structure function shows a vanishing finite-time
asymmetry which is likely to be universal and controlled by cubic
corrections of the NLFH theory. The distribution of time-integrated
currents do not show any indication for the relevance of cubic or
higher corrections. Overall, we have found strong indications that the
NLFH theory works properly for systems discrete in space {\em and}
time (i.e. parallel update). Relevant quantities determining the asymptotic
behavior are the current-density relation and the compressibility.

The slow relaxation modes are controlled by the universality class of
the system. Monte Carlo simulations showed that, observables recorded
in systems with insuffciently relaxed initial states show strong
deviations to observables recorded in stationary systems. To overcome
effects caused by insufficient relaxation, we have derived a
relaxation criterion (\ref{eq:relax_time}) for single species models
that exhibit a nonlinear current-density relation. This criterion
yields a precise estimate for the minimal time necessary to reach a
state that can be considered stationary in simulations. It would be of
interest to understand better the survival of universality in
transient regimes of the NaSch model with time-dependent boundaries.

One expects NLFH to hold as well for multi-species models such as
traffic models that incorporate cars as well as buses. In this case
one expects fluctuations in the eigenmodes of NLFH to be described by
explicitly known universal scaling functions, see
\cite{Popk15b,Popk16} for the general multi-species case.  In
\cite{CdGHS} a more mathematical treatment of KPZ modes has been given
for two-species models where NLFH is not postulated and universal
distributions have been derived from first principles, confirming NLFH
predictions.

 \begin{acknowledgments}
   JdG would like to thank Tim Garoni for discussions and gratefully
   acknowledges financial support from the Australian Research Council.
  JS thanks the University of Melbourne, where parts of this work were
  done, for hospitality and support.  He acknowledges ACEMS and the
  Bonn-Cologne-Graduate-School for covering travel expenses.  This
  work was supported by Deutsche Forschungsgemeinschaft (DFG) under
  grant SCHA 636/8-2. AS, JS and GS acknowledge support by the German
  Excellence Initiative through the University of Cologne Forum
  "Classical and Quantum Dynamics of Interacting Particle Systems".
\end{acknowledgments}


\begin{appendix}
\section{Simulation method}
In order to run efficient  Monte Carlo simulations, it is recommendable
to utilize  translational invariance  due to periodic boundary
conditions and stationarity allowing for ergodic measurements
by averaging  over space and time, i.e.
\begin{eqnarray}
\label{eq:MonteCarloEstimator}
&&\tilde{f}_{M,\tau}\left[\{n_{x,t}\}_L\right]( \vec{x}_*,\vec{t}_* )=\nonumber\\
  &&\quad\frac{1}{LM}\sum_{l,m=1}^{L,M} f\left[\{n_{x,t}\}_L\right](\vec{x}_*
  +l\vec{1},\vec{t}_*+m\tau\vec{1})
\end{eqnarray}
where $\tilde{f}$ and $f$ are Metropolis-Hastings Monte-Carlo
estimators evaluating a single stationary Markov Chain $\{n_{x,t}\}_L$
of a system with $L$ sites. The evaluation points of interest are the
positions $\vec{x}_*$ and times $\vec{t}_*$, their corresponding ones
vectors $\vec{1}=(1,\ldots,1)^t$ shift these points in order to make
use of the translational invariance and stationarity.  The average of
$\tilde{f}_{M,\tau}$ over $P\rightarrow\infty$ independent
realisations $\{n_{x,t}\}_L$ guarantees the convergence to the desired
quantity

\begin{eqnarray}
\label{eq:MonteCarloEstimator2}
&&\mathbb{E}\left(\tilde{f}_{M,\tau}(\vec{x}_*,\vec{t}_*)\right)=\nonumber\\
&&\lim_{P\rightarrow\infty}\frac{1}{P}\sum_{p=1}^{P} \tilde{f}_{M,\tau}\left[\{n_{x,t}\}_{L,p}\right](\vec{x}_*,\vec{t}_*)
\end{eqnarray}
Note that, in case of stationarity and translational invariance one
has
$\mathbb{E}\left(\tilde{f}_{M,\tau}(\vec{x}_*,\vec{t}_*)\right)=\mathbb{E}\left(f(\vec{x}_*,\vec{t}_*)\right)$,
whereas $\tilde{f}_{M,\tau}$ supports a significantly lower variance
than $f$ and therefore consumes less computation time to reach the
desired accuracy.  Further, the time between two ergodic measures
$\tau$ may serve as a variance reduction parameter allowing to
minimize the uncertainty of the estimator $\tilde{f}_{M(\tau),\tau}$
under fixed computation cost.

E.g. the estimator $\tilde{f}_{M,\tau}$ for the single lane dynamical
structure function $S(x,t)=\mathbb{E} (\tilde{f}_{M,\tau} )$ (see
Eq.~(\ref{eq:SingleLaneStrucFCT})) is based on
$f\left[\{n_{x,t}\}_L\right]((0,x)^t,(0,t)^t)=n_{0,0}n_{x,t}-\rho^2$.

Independent stationary Markov Chains $\{n_{x,t}\}$ are realised by
using independent initial states $\{n_{x,0}\}$ drawn from stationary
distribution, and propagated according to the systems update rules
with independent sets of random numbers.  In case of unknown
stationary distribution ($v_\mathrm{max}>1$), we use the initial
configuration where all vehicles are equally distributed and assigned
to their maximum velocity. In order to reach the stationary limit,
each configuration is independently propagated with at least
$T_{\text{relax}}$ updates (see Eq.~(\ref{eq:relax_time})).

All pseudo random numbers throughout this paper are generated by the Mersenne
Twister generator,  implemented in the \texttt{C++} standard library
random.  

\end{appendix}



\end{document}